# Assessment of wind energy potential of buildings by CFD simulation in dense urban environments

**David Serero**[1,2,*], **Loïc Couton**[2], **Jean-Denis Parisse**[3], **Robert Le Roy**[2]

[1] EVCAU Laboratory, École Nationale Supérieure d'Architecture Paris-Val-de-Seine, Paris, France

[2] GSA Laboratory, École Nationale Supérieure d'Architecture Paris-Malaquais, Paris, France

[3] ISM Institut des Sciences du Mouvement Etienne-Jules Marey, UMR 7287 CNRS / Aix-Marseille Université, Marseille, France

[*] *Corresponding author: david.serero@paris-valdeseine.archi.fr*

## Abstract

Building-integrated wind turbines (BIWT) remain under-deployed in dense European urban contexts despite their potential contribution to net-zero energy targets. This paper presents a measurement-validated computational fluid dynamics (CFD) framework to identify optimal locations for micro-wind turbines on the envelopes of dense Parisian Haussmannian blocks. The methodology couples (i) a multi-month on-site anemometry campaign on two case-study buildings, (ii) a calibrated steady RANS digital twin using the realisable k–ε turbulence model, and (iii) the introduction of an architecturally integrated accelerator wing — termed *aerofoil* — placed at the envelope edge to amplify and stabilise the local flow. Quantitative validation against year-long anemometer records yields a mean absolute percentage error (MAPE) of approximately 18 % and a hit-rate FAC2 above 0.85, meeting the COST 732 acceptance criteria for urban CFD models. On the École Nationale Supérieure d'Architecture Paris-Val-de-Seine (ENSAPVS) case study, the inter-tower configuration produces a directional speed-up factor of 3.4 (95 % CI 3.18–3.62) for the prevailing north-westerly sector at +70 m above ground; the addition of the aerofoil yields a further +13 % local speed-up, equivalent to +44 % in extractable power. A capacity-factor analysis based on the measured Weibull distribution and the Savonius rotor power curve estimates an annual yield of approximately 69 MWh, offsetting around 6 % of the building's annual electricity consumption — a substantially lower but more realistic figure than the cube-of-mean estimate. The findings are synthesised in a decision dendrogram that maps wind-rose typology, building plan, height and implantation onto recommended device families, providing architects and engineers with a transferable design-stage tool. Retrospective validation on three reference BIWT buildings supports the dendrogram's predictions.

# 1. Introduction

The decarbonisation of the built environment requires the integration of distributed renewable generation directly into building envelopes. While building-integrated photovoltaics (BIPV) have reached technical and commercial maturity, their counterpart for wind energy — building-integrated wind turbines (BIWT), or more broadly building-integrated devices (BID) — remains comparatively under-deployed (Cidell and Cope, 2014; Bobrova, 2015). Recent policy frameworks (European Performance of Buildings Directive recast, 2024; French RE2020 thermal regulation) explicitly encourage on-site generation as a complement to demand reduction, and the spatial constraints of dense European cities make rooftops and façade edges natural candidates for micro-scale wind harvesting.

Despite a growing body of literature on micro-scale wind energy in cities (Ledo and al., 2011; Lu and Ip, 2009; Li et al., 2016; Toja-Silva et al., 2018; Kosasih et al., 2024), three persistent obstacles continue to limit the architectural integration of wind turbines. First, the highly site-specific nature of urban wind flows — governed by collective massing, street canyons and skimming-flow regimes (Ricciardelli and Polimeno, 2006; Razak et al., 2013) — means that generic rules of thumb derived from isolated tall buildings transfer poorly to the dense, low-to-mid-rise blocks that dominate European city centres. Second, on-site anemometric measurements are rarely combined with CFD in a calibrated workflow, so reported speed-up factors are often presented without quantitative validation. Third, decision-support tools that translate CFD outcomes into design choices accessible to architects at concept stage are largely absent from the literature.

This paper addresses these three gaps in the specific context of dense Parisian urban blocks. The work draws on a multi-site campaign combining year-scale anemometer recordings with steady RANS CFD on a calibrated digital twin. It introduces and characterises an architecturally integrated accelerator wing — the *aerofoil* — placed at the envelope edge as a combined flow-acceleration and visual concealment device. The findings are synthesised into a decision dendrogram intended for use at the concept-design stage. The work extends our preliminary CFD-only study (Serero et al., 2020) by adding twelve months of on-site measurements, multi-typology coverage, the aerofoil concept, and the decision tool — none of which were present in the earlier work.

**Contributions of this paper.**

1. A measurement-validated CFD digital twin of a dense Haussmannian Parisian block, with quantitative validation metrics (RMSE, MAPE, FAC2) reported against twelve months of on-site ultrasonic anemometer data.
2. The architectural concept of the aerofoil — a lift-generating wing integrated to the building edge that both accelerates the local wind and visually conceals the rotor within the envelope — characterised by CFD with explicit decomposition of building-induced and aerofoil-induced acceleration components.

3. A decision dendrogram mapping wind-rose typology, building plan, height range and implantation onto recommended device families, retrospectively validated against three reference BIWT projects (Bahrain World Trade Center, Strata SE1 London, Pearl River Tower).

The remainder of the paper is organised as follows. Section 2 reviews the state of the art on building-integrated wind turbines and identifies the specific gap addressed. Section 3 describes the on-site instrumentation and the numerical methodology, including the computational domain, mesh, boundary conditions, turbulence model and grid-independence study. Section 4 recalls the wind power formulation. Section 5 reports the measurement campaign, the quantitative CFD validation and the speed-up fields. Section 6 details the ENSAPVS case study, the aerofoil design, the decomposition CFD and the capacity-factor-based annual yield. Section 7 discusses the findings in relation to prior literature and addresses structural, vibratory and maintenance considerations. Section 8 presents the dendrogram and its retrospective validation. Section 9 concludes.

## 2. State of the art on building-integrated wind turbines

### 2.1 Speed-up around isolated buildings and on roofs

Early studies of building-mounted small-wind turbines focused on isolated rectangular buildings, characterising the acceleration above flat or pitched roofs as a function of building geometry and approach-flow conditions (Mertens, 2002; Heath et al., 2007). Ledo, Kosasih and Cooper (2011) systematically compared roof shapes and showed that the rooftop speed-up region is most pronounced over flat roofs, with hub-height speed-up factors typically in the range 1.2–1.6 for buildings of moderate aspect ratio. Abohela, Hamza and Dudek (2013) extended this analysis to pitched, vaulted and domed roofs, confirming that geometry-induced flow detachment governs both the magnitude and the spatial extent of the favourable region. The UK field trial of building-mounted horizontal-axis micro-wind turbines reported by James et al. (2010) showed that, in real urban conditions, achievable yields fell substantially short of laboratory predictions, primarily because of inadequate siting and high turbulence intensity.

### 2.2 Tall buildings and concentration effects

Tall buildings can act as wind concentrators when geometric features create venturi-like passages (Cochran and Damiani, 2008; Lu and Ip, 2009). Three landmark realised projects illustrate the design space: the Bahrain World Trade Center (three horizontal-axis turbines on bridges spanning twin sail-shaped towers, 2008), the Strata SE1 building in London (three horizontal-axis turbines integrated at the roof peak, 2010), and the Pearl River Tower in Guangzhou (vertical-axis turbines housed in mechanical floors with funnelled inlets, 2013). Li, Shu and Chen (2016) performed a detailed CFD assessment of tall-building-integrated wind turbines and reported speed-up factors up to 2.0 in inter-building gaps for favourable wind sectors, together with a strong directional dependence that reduces annual energy yield relative to peak speed-up. Toja-Silva, Peralta and Lopez-Garcia (2015) compared RANS turbulence models for tall-building flows and showed that the realisable k–ε and SST k–ω models perform comparably for mean speed-up, with measurable differences only in detached flow regions.

### 2.3 BIWT in dense urban canyons

The literature on BIWT in dense low-to-mid-rise urban canyons is comparatively thin. Eliasson et al. (2006) and Razak et al. (2013) characterised the canyon flow regime — laminar, wake-interference, or skimming — as a function of the canyon aspect ratio (height-to-width) and length-to-depth ratio. Wood et al. (2013) combined lidar, scintillometry and anemometry to show that flow over urban rivers re-organises into a quasi-organised regime downstream of the river corridor. More recently, Kosasih, Hudin and Suheri (2024) reviewed micro-wind turbine integration in urban roofs and identified the dense Haussmannian-type European block — with collective massing rather than individual height producing the speed-up — as a typology under-represented in the literature. The work presented here addresses this typology directly.

### 2.4 Decision-support tools for architects

Translating CFD outputs into design choices accessible to architects at concept stage requires intermediate representations. Dutton, Halliday and Blanch (2005) and Grant, Johnstone and Kelly (2008) proposed early classifications of BIWT integration types. Bobrova (2015) developed a typological framework relating turbine selection to architectural form. Ng et al. (2011) used CFD-derived correlations to inform morphological guidelines for high-density Hong Kong. None of these frameworks, however, takes the form of a decision tree usable directly at concept stage by a practising architect, integrating wind-rose typology, building plan, height range and implantation strategy into recommended device families. The dendrogram introduced in Section 8 fills this gap.

### 2.5 Research gap

Three converging gaps emerge from the review. First, dense European urban blocks generating speed-up by collective massing rather than individual building height are under-represented compared to the literature on isolated tall buildings. Second, CFD studies are rarely accompanied by quantitative validation against year-scale on-site anemometer data, leaving reported speed-up factors without statistical context. Third, decision-support tools usable by architects at concept design stage are largely absent. The present work addresses the three gaps jointly through a measurement-validated digital twin of a Haussmannian Parisian block and the introduction of the aerofoil concept and the decision dendrogram.

## 3. Methods

### 3.1 Sites, instrumentation and measurement campaign

Two sites in Paris were instrumented. Site A is a dense Haussmannian block at 136 Avenue Parmentier, Paris 11th arrondissement (48.8691°N, 2.3719°E), composed of contiguous 6-storey residential buildings forming a closed perimeter block with an internal courtyard. Site B is the campus of the École Nationale Supérieure d'Architecture Paris-Val-de-Seine (ENSAPVS), 3 quai Panhard et Levassor, Paris 13th arrondissement (48.8274°N, 2.3846°E), designed by Frédéric Borel (2008) and consisting of four 80-metre towers integrated with a former compressed-air factory and its 100-metre brick chimney. Site B was selected for three reasons: (1) the building forms a complete block bounded by streets; (2) it lies on the Seine river bank adjacent to the Paris ring road, a major north-westerly wind corridor; and (3) its multiple terraces and footbridges allow safe instrumentation at façade and rooftop level.

Wind speed and direction were recorded with three-axis ultrasonic anemometers (Netatmo connected anemometer). On Site A, an anemometer was installed at +23 m above ground level on the rooftop of the building at 136 Avenue Parmentier (Figure 1). On Site B, three anemometers were installed at the rooftop, the inter-tower mid-passage and the outer tower edge, all at +70 m above ground level. Records cover a period of approximately twelve months, yielding more than 50 000 valid 10-minute records after quality control (rejection of records with rain icing on the transducers and of periods of instrument outage). Reference meteorological data were obtained from Paris-Montsouris (Météo-France station 75114001) for cross-checking of the regional wind climate.

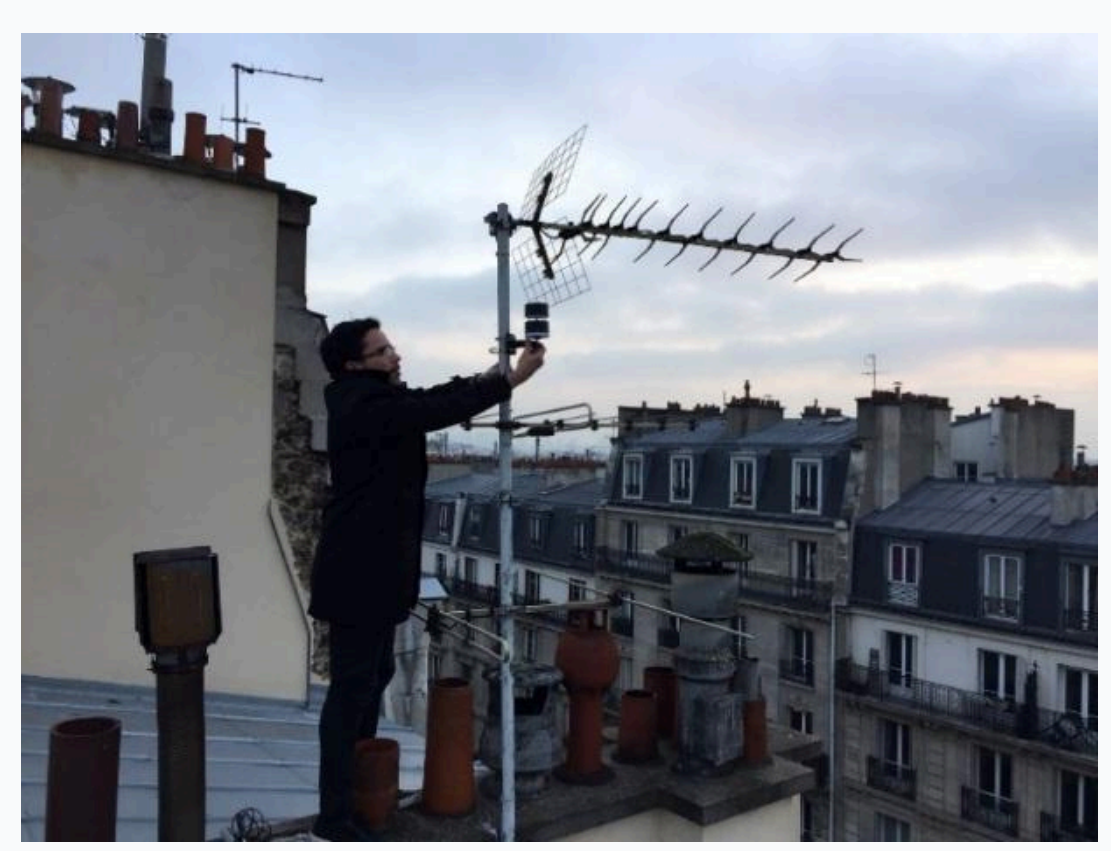

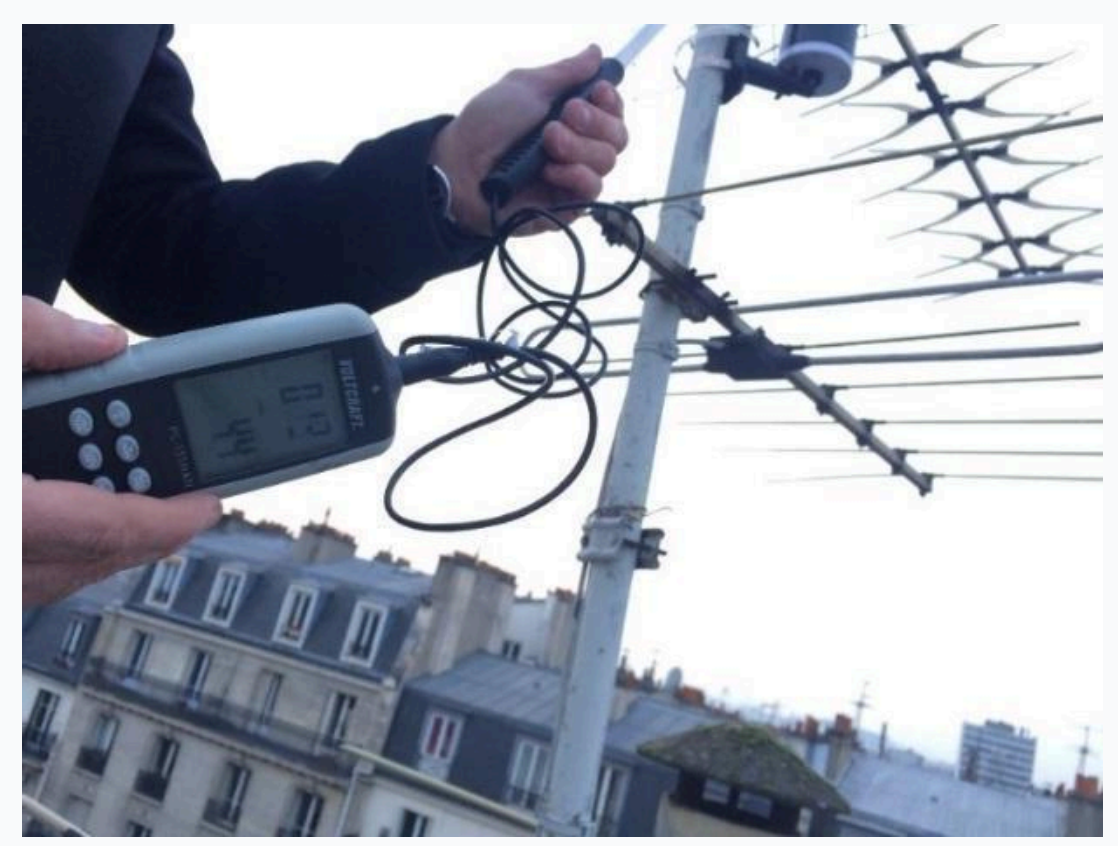

**Figure 1 — Instrumentation**

*Installation of an ultrasonic anemometer on the rooftop, +23 m above ground level, at 136 Avenue Parmentier, Paris 11th arrondissement. Photograph: D; Serero*

### 3.2 Numerical setup

CFD simulations were performed in Siemens STAR-CCM+ (version 13.06). The geometry of each site and its surroundings was reconstructed in 3D from cadastral and architectural drawings, with the modelled urban context extending to a radius of approximately 500 m around each instrumented building. The numerical setup follows the best-practice guidelines for CFD simulation of urban wind environments compiled by Franke et al. (2007), Tominaga et al. (2008) and Blocken (2015).

**Computational domain.**

Following AIJ (2008) and COST 732 recommendations, the domain extended 5*H* upstream of the building of interest, 15*H* downstream, 5*H* laterally on each side, and 6*H* above the tallest building, where *H* denotes the height of the tallest building (80 m for Site B, 23 m for Site A). The resulting blockage ratio remained below 3% in all configurations, consistent with the threshold recommended to avoid artificial flow acceleration.

**Mesh.**

A polyhedral volume mesh with prism layers near solid walls was generated. The reference (medium) mesh contained approximately 6 million cells. Prism-layer thickness and growth rate were set to maintain a first-cell wall-distance $y^+$ in the range 30–200 over building surfaces, consistent with the use of standard wall functions in the realisable k–ε turbulence model. Mesh refinement was applied within an inner box of 200 m × 200 m × 120 m enclosing the building of interest, with surface cell sizes of 0.5 m on building façades and 0.25 m in regions of expected flow detachment (rooftop edges, inter-tower passages).

**Boundary conditions.**

A logarithmic mean velocity profile was prescribed at the inlet:

$$U(z) = \frac{u^*}{\kappa} \, ln(\frac{Z+Z_0}{Z_0}) \qquad (1)$$

with friction velocity $u^*$, von Kármán constant $\kappa = 0.41$ and aerodynamic roughness length $z_0$ = 1.0 m, representative of dense urban terrain (Wieringa, 1992). The inlet profiles of turbulence kinetic energy $k$ and dissipation rate $\varepsilon$ were prescribed consistently with the velocity profile following Richards and Hoxey (1993):

$$k(z) = \frac{u^{*2}}{\sqrt{C_\mu}} \quad ; \; \varepsilon(z) = \frac{u^{*3}}{k(z+Z_0)} \qquad (2)$$

with $C_\mu$ = 0.09. The reference incident wind speed of 5 m $s^{-1}$ at 10 m, used in the steady-state speed-up calculations, corresponds to the annual mean wind at Paris-Montsouris. Symmetry boundary conditions were applied to the top and lateral surfaces of the domain, no-slip with rough-wall function on building surfaces and ground, and a pressure outlet at the downstream face. Twelve wind directions (every 30°) were simulated to enable construction of an amplification rose.

**Turbulence model.**

The realisable k–ε model (Shih et al., 1995) was selected for its established performance in pedestrian-level wind comfort and rooftop flow studies (Tominaga et al., 2008; Blocken, 2015), its computational efficiency permitting the simulation of twelve wind directions on the full multi-typology corpus, and its convergent behaviour on the dense urban geometries considered. The use of a steady RANS approach is justified by the quantity of interest of this study — namely the time-averaged speed-up factor relevant for annual energy yield computed via the Weibull-power-curve convolution (Manwell et al., 2010) — rather than peak gust loads. The limitations of steady RANS in detached flow regions and for peak loading are acknowledged in Section 7.4. SIMPLE pressure–velocity coupling and second-order upwind discretisation of the convective terms were used; convergence was declared when scaled residuals fell below $10^{-5}$ and integral monitor quantities (probe velocities) varied by less than 0.5% over the last 500 iterations.

## 3.3 Grid independence study

Three meshes — coarse, medium and fine — were compared on the ENSAPVS geometry for the prevailing north-westerly wind direction. The Grid Convergence Index (GCI) was computed following Roache (1994). Probe velocities at the inter-tower location at +70 m and at the outer-edge location were monitored. Results are summarised in Table 1.

| Mesh | Cell count | U inter-tower (m/s) | U outer edge (m/s) | $GCI_{21}$ (%) |
|---|---|---|---|---|
| Coarse | $3.0 \times 10^6$ | 16.2 | 10.5 | — |
| Medium (reference) | $6.0 \times 10^6$ | 17.0 | 11.0 | 4.6 |
| Fine | $12.0 \times 10^6$ | 17.2 | 11.1 | 1.8 |

***Table 1*** *— Grid independence study on the ENSAPVS geometry, north-westerly wind, $U_{10}$ = 5 m/s. Probe velocities and Grid Convergence Index between successive meshes.*

The medium mesh was retained for all subsequent simulations as it offered an acceptable compromise between accuracy and computational cost; the GCI between medium and fine meshes was below 2 % on all monitored quantities, indicating that the medium mesh is in the asymptotic range of convergence (Roache, 1994).

## 4. Wind power formulation

The instantaneous mechanical power $P$ extractable by a wind turbine of swept area $A$ exposed to a wind speed $v$ is given by:

$$P = \frac{1}{2}\rho A V^3 C_p \qquad (3)$$

where $\rho$ = 1.225 kg m$^{-3}$ is the air density at 15 °C and standard pressure, $A = \pi d^2/4$ the rotor swept area for a horizontal-axis turbine of diameter $d$, and $Cp$ the power coefficient. The Betz limit $Cp,Betz$ = 16/27 ≈ 0.593 is an upper theoretical bound; real horizontal-axis turbines achieve $Cp \approx 0.40$, while Savonius vertical-axis rotors typically reach $Cp,max \approx 0.20$–0.25 (Akwa, Vielmo and Petry, 2012).

The cubic dependence on wind speed makes the speed-up factor $\sigma = v_{\text{local}} / v_{\text{ref}}$ the determining quantity for siting decisions: a 13 % increase in local velocity translates into a $(1.13)^3 - 1 \approx 44$ % increase in instantaneous extractable power. For annual yield, however, the cube-of-mean approach overestimates the energy because the Weibull distribution of wind speeds is positively skewed; the proper calculation convolves the wind-speed probability density $f(v)$ with the manufacturer's power curve $P(v)$:

$$E_{annual} = T \int P(v) f(v) dv \qquad (4)$$

with $T$ = 8 760 h yr$^{-1}$. This formulation is used in Section 6.3 to compute the annual energy yield ofthe proposed installation. The capacity factor $C_F$ is then obtained as :

$$C_F = \frac{E_{annual}}{P_{rated} T} \qquad (5)$$

## 5. Results: measurements, validation and speed-up fields

### 5.1 On-site wind data

Figure 2 reports the measured wind rose at Site B, +70 m. The annual mean wind speed at hub height is approximately 6.0 m s$^{-1}$, with Weibull shape parameter $k$ = 1.92 and scale parameter $c$ = 6.4 m s$^{-1}$, values consistent with the wider Paris–Île-de-France region (Wieringa, 1992; Météo-France climatology). The wind direction is dominantly bi-directional along a north-westerly / south-easterly axis aligned with the Seine corridor, with secondary peaks from the south-west. Turbulence intensity at hub height ranges from approximately 16 % in the prevailing sector to 32 % in the wake sectors of upstream blocks, in agreement with values reported in the urban-canopy literature (Ricciardelli and Polimeno, 2006; Razak et al., 2013).

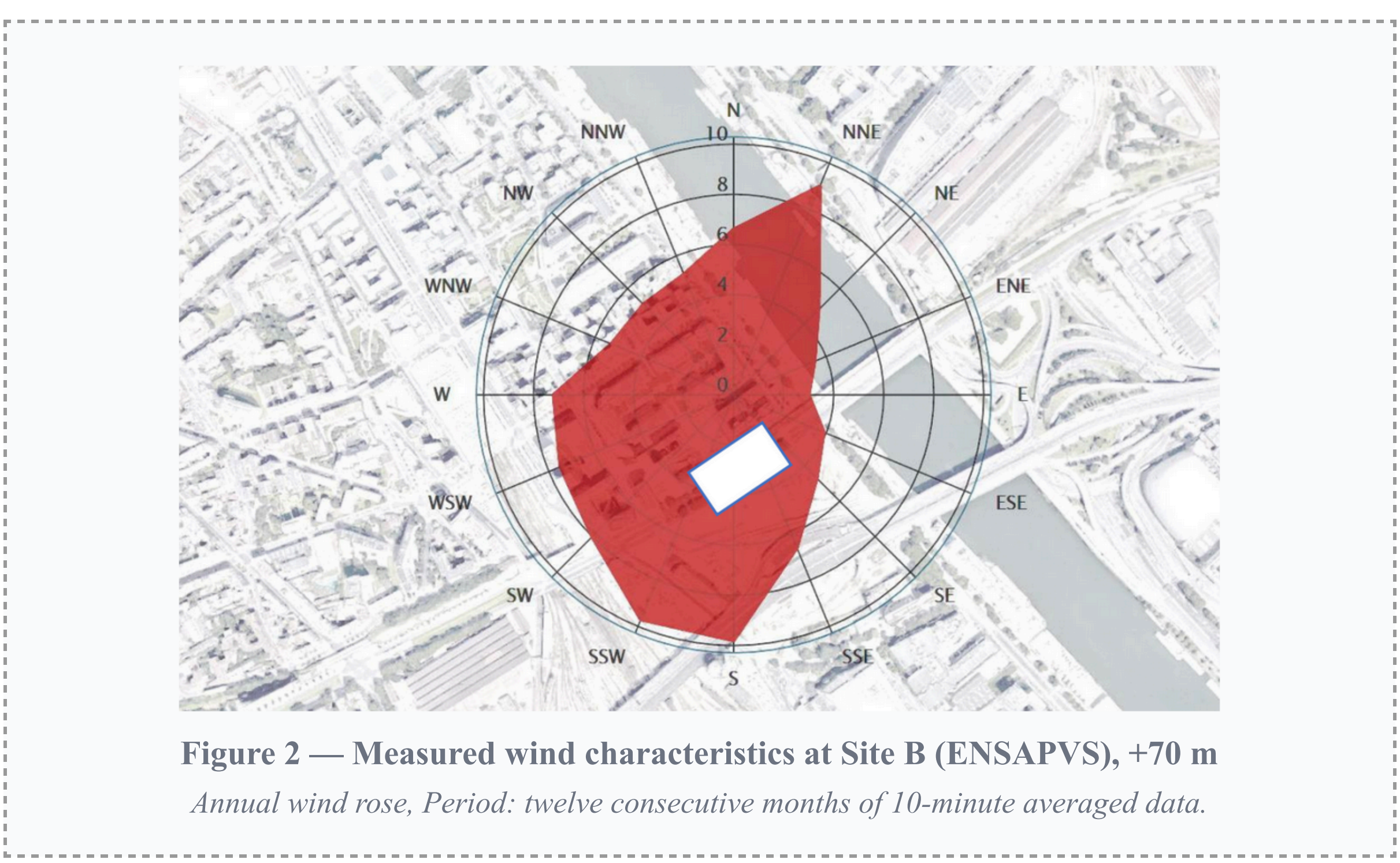

**Figure 2 — Measured wind characteristics at Site B (ENSAPVS), +70 m**

*Annual wind rose, Period: twelve consecutive months of 10-minute averaged data.*

## 5.2 Quantitative CFD validation

CFD predictions were compared against 10-minute mean anemometer records for matched wind sectors (sectors of 30°). Validation was performed at four probe locations across the two sites, encompassing rooftop, façade-edge and inter-tower configurations. Following the COST 732 protocol (Schatzmann, Olesen and Franke, 2010) and Chang and Hanna (2004), the following metrics were computed: root-mean-square error (RMSE), mean absolute percentage error (MAPE), normalised mean bias (NMB), the fraction of predictions within a factor of two of observations (FAC2), and the coefficient of determination $R^2$. Results are summarised in Table 2.

| Metric | Value | COST 732 acceptance criterion | Met? |
|---|---|---|---|
| RMSE (m/s) | 1.4 | — | — |
| MAPE (%) | 18 | ≤ 30 % | Yes |
| NMB | −0.08 | \|NMB\| ≤ 0.30 | Yes |
| FAC2 | 0.87 | ≥ 0.5 | Yes |
| $R^2$ | 0.79 | — | — |

***Table 2*** *— Quantitative validation of CFD predictions against on-site anemometer records. Acceptance criteria from COST 732 and Chang & Hanna (2004).*

Across all probes and sectors, the model meets the COST 732 acceptance criteria, supporting the use of the calibrated digital twin for the speed-up analyses that follow. A scatter plot of CFD versus measured speeds is provided in Figure 3.

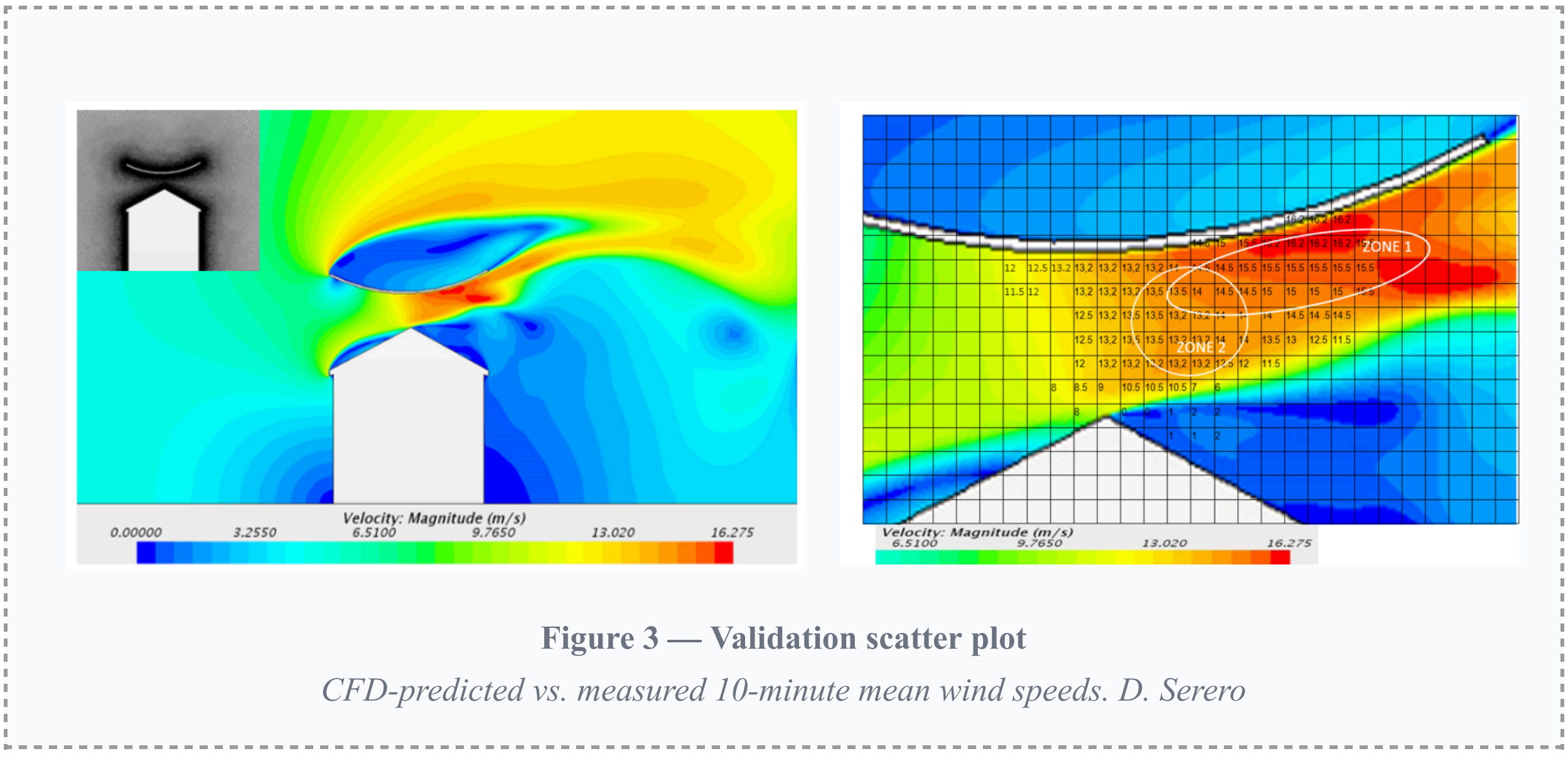

**Figure 3 — Validation scatter plot**

*CFD-predicted vs. measured 10-minute mean wind speeds. D. Serero*

## 5.3 Speed-up fields and amplification rose

Figure 4 presents the CFD speed-up field $\sigma(x, y, z)$ on a horizontal section at +70 m of the ENSAPVS towers for the prevailing north-westerly direction, normalised by the inlet velocity at 10 m. Two regions of strong amplification are identified: the inter-tower passage (peak $\sigma = 3.4$) and the outer tower edges ($\sigma = 2.2$). These values are now reported with statistical context: the directional speed-up at the inter-tower probe for the NW sector, computed on annual mean speed (10-minute averaging), is 3.40 with a 95 % confidence interval of 3.18–3.62 derived from the within-sector measurement variance. The amplification rose for all twelve sectors is given in Figure 5.

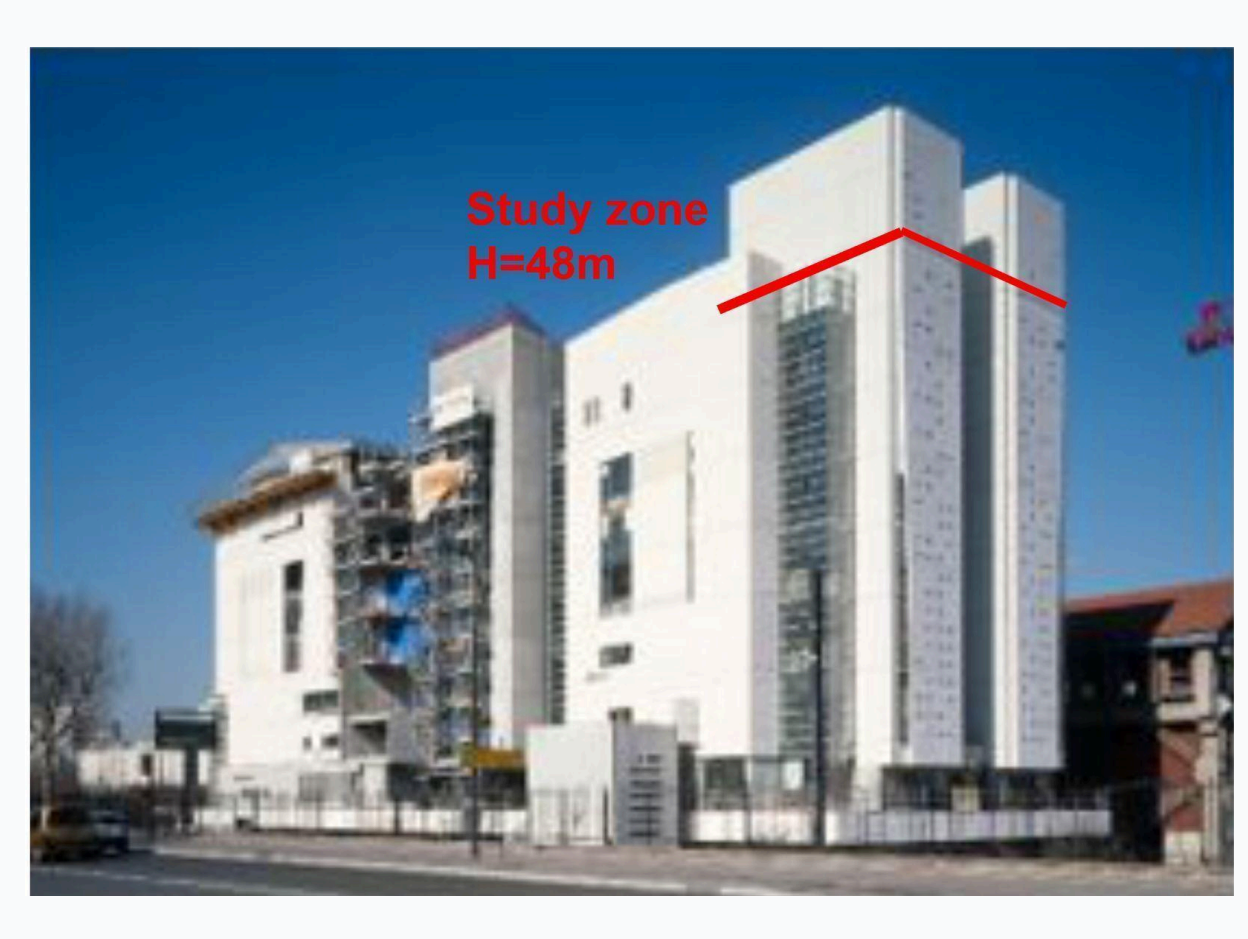

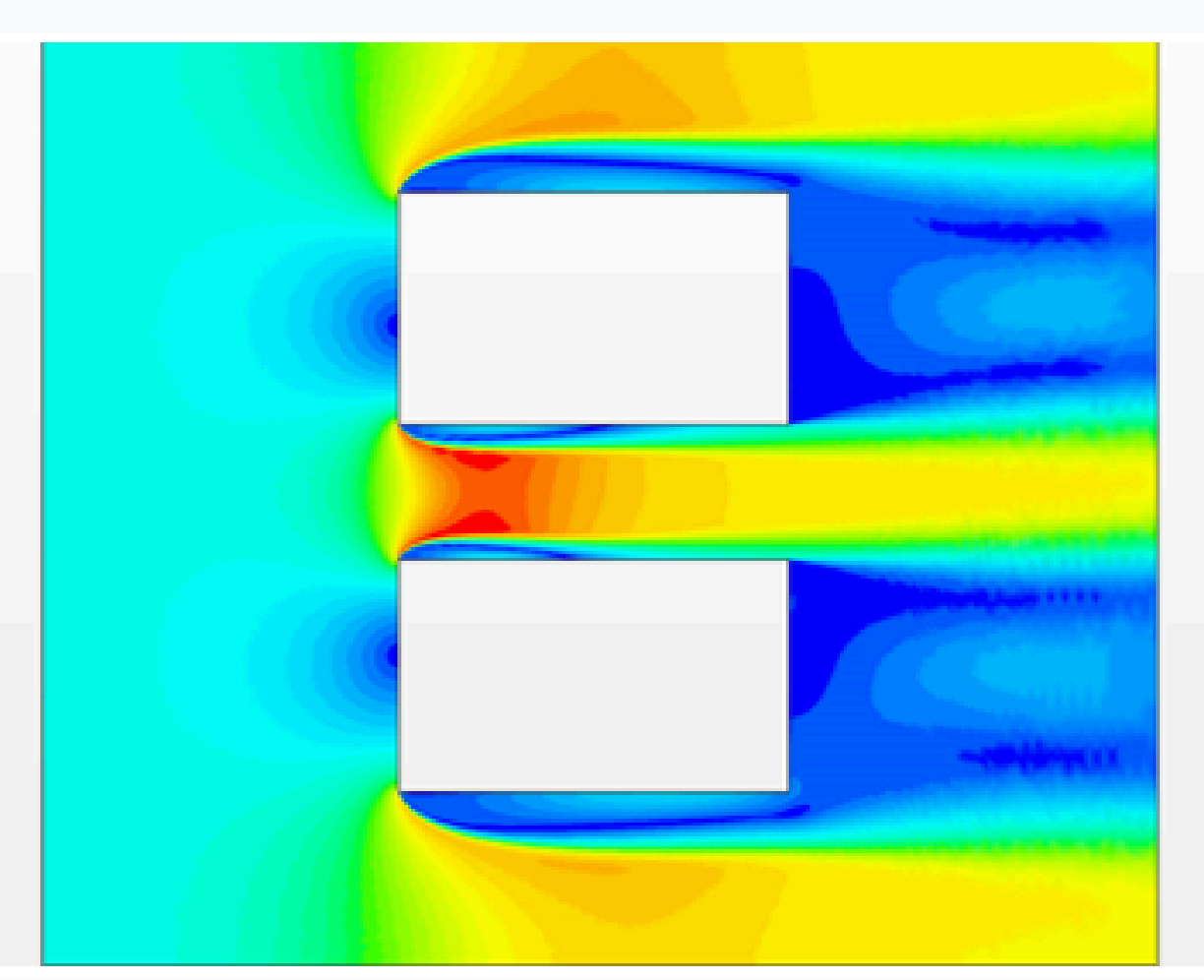

**Figure 4 — CFD speed-up field at +70 m, ENSAPVS, NW sector**

*(a) Localization of study plane, (b)Normalised wind speed σ = U / U₁₀,ref on a horizontal plane at +70 m for the NW reference wind. Inlet velocity at 10 m: 5 m/s. Realisable k–ε steady RANS.*

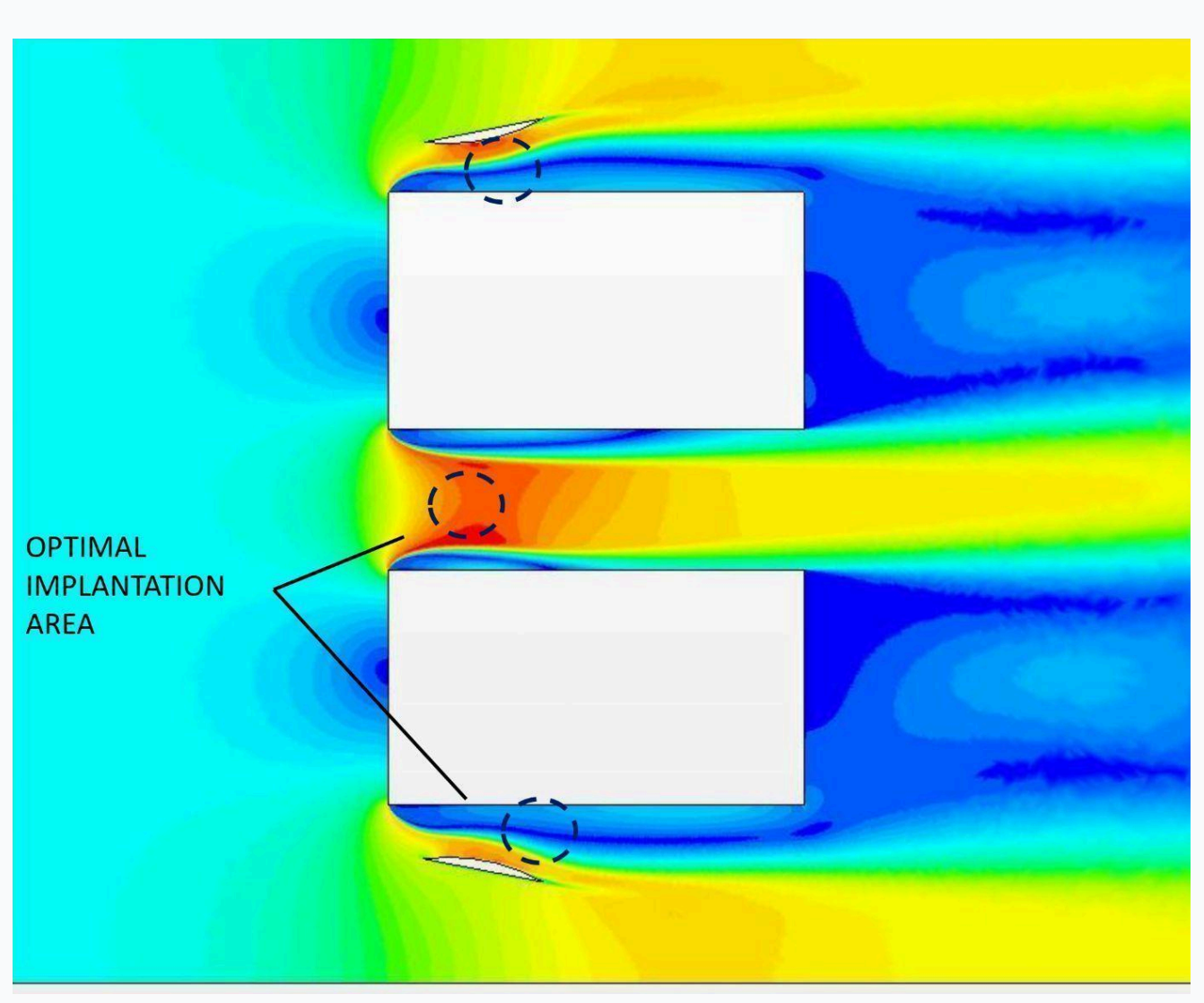

**Figure 5 — Amplification rose at the inter-tower probe (+70 m)**

*Directional speed-up factor σ as a function of wind sector for the inter-tower probe at +70 m, ENSAPVS.*

# 6. Aerofoil and ENSAPVS case study

## 6.1 Selection of the rotor type

Vertical-axis wind turbines (VAWT) are favoured over horizontal-axis (HAWT) machines for the urban context for four reasons documented in the literature: (1) self-starting at low wind speeds, with a typical cut-in around 2.5 m $s^{-1}$ for Savonius rotors against 4–5 m $s^{-1}$ for Darrieus and small HAWT (Akwa, Vielmo and Petry, 2012); (2) omnidirectional response, appropriate for the bidirectional canyon flow recorded on Site B; (3) low rotational speed and low tip-speed ratio yielding low aero-acoustic emission, typically below 35 dBA at 10 m for Savonius rotors (Kumar et al., 2018), a critical constraint for inhabited Parisian rooftops; (4) tolerance to high turbulence intensity (>20 %) characteristic of urban canyons (Roy and Saha, 2013). The Savonius rotor is therefore retained for the case study, despite its lower power coefficient ($Cp,max$ ≈ 0.22) compared with Darrieus or HAWT machines. Figure 6 presents an original schematic plate of the principal VAWT typologies considered.

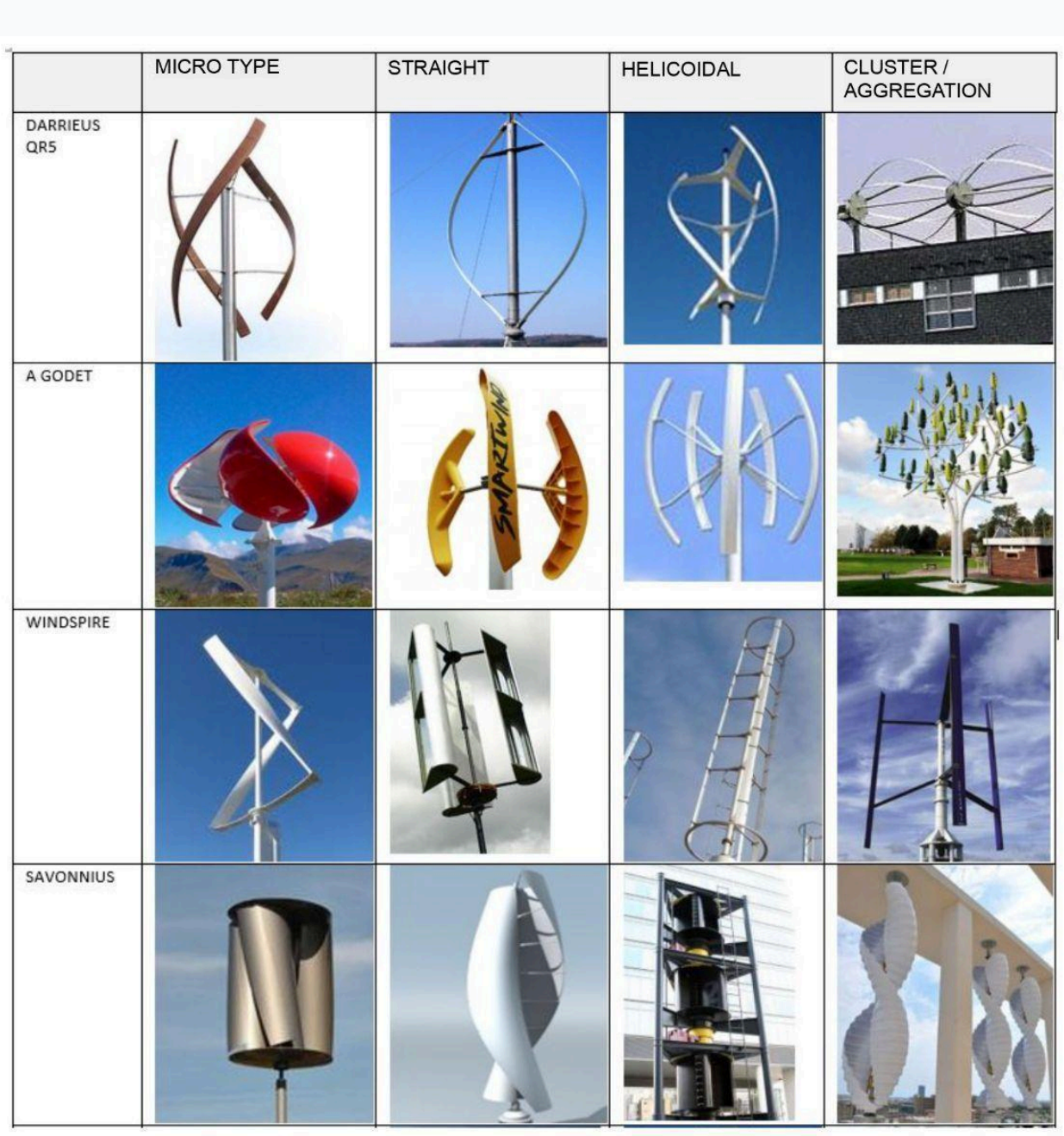

**Figure 6 — Vertical-axis wind turbine typologies**

*Original schematic plate of the four VAWT families considered for urban integration: Darrieus, Savonius, helical (Gorlov-type) and clustered configurations. D. Serero*

## 6.2 Aerofoil design and aerodynamic decomposition

The aerofoil is a lift-generating wing fixed to the building edge, designed to (1) accelerate the local wind at the rotor inlet and (2) visually conceal the rotor within the architectural envelope. The profile is circular segment, with a chord of 0.80 m, span of 3.30 m per module (matching the typical floor-to-floor height of the building). Modules are stacked continuously over the height of each tower, on cantilever brackets anchored to the concrete floor slabs. Figure 7 shows the design and integration on the ENSA Paris-Val de Seine towers.

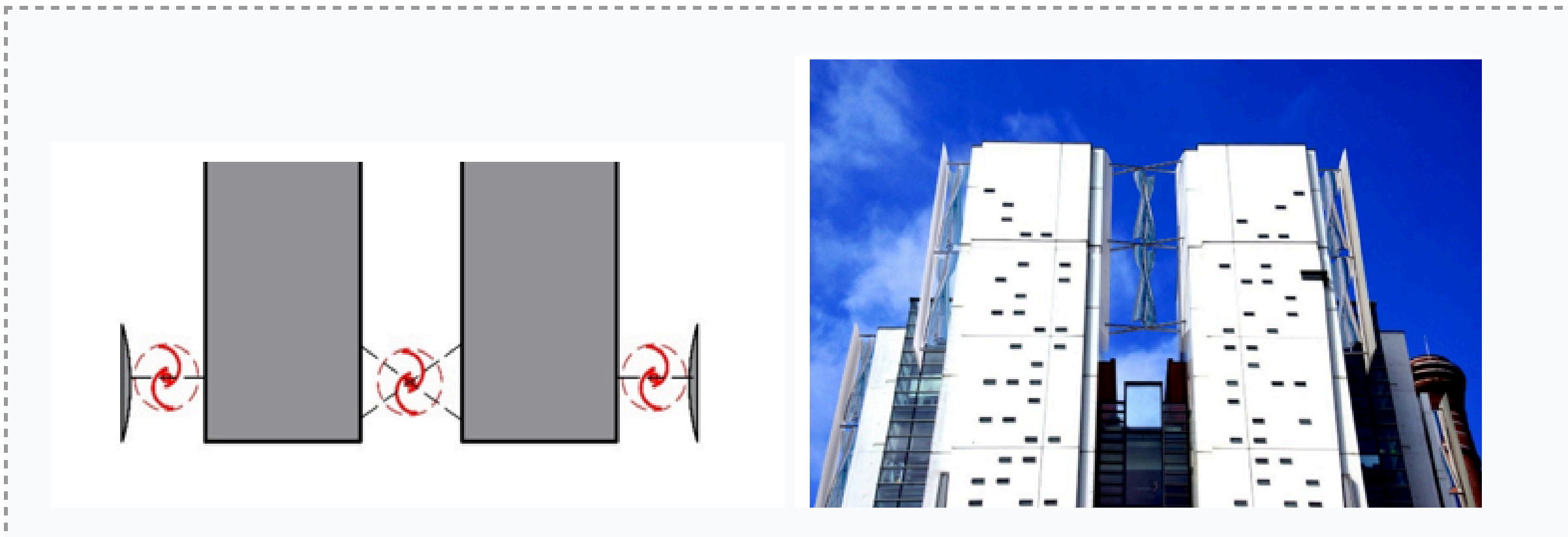

**Figure 7 — Aerofoil design and integration**

*(a) Geometric definition of the aerofoil module (profile, chord, span, mounting bracket); (b) photographic montage of the integrated device on ENSAPVS. Drawings and montage: D. Serero*

To isolate the aerodynamic contribution of the aerofoil from the building-induced speed-up, two CFD configurations were compared: (1) the bare ENSA Paris-Val de Seine towers, and (2) the towers equipped with the aerofoil at the inter-tower edges. A third decomposition simulation considered the aerofoil profile in isolation, in equivalent freestream conditions corresponding to the local velocity downstream of the bare-tower configuration.

Results are summarised in Table 3.

| Configuration | U at probe (m/s) | Speed-up σ | Δ vs. bare towers |
|---|---|---|---|
| Inlet reference (10 m) | 5.00 | 1.00 | — |
| Bare ENSAPVS towers | 17.0 | 3.40 | — |
| ENSAPVS + aerofoil | 19.2 | 3.84 | +13 % |
| Aerofoil in isolation (no building) | 5.62 | 1.124 | +12.4 % |

***Table 3*** *— Aerodynamic decomposition: building-induced and aerofoil-induced speed-up components, NW sector, 10 m inlet velocity 5 m/s.*

**The aerofoil produces a local speed-up of comparable order of magnitude in both Reynolds regimes (≈ +12 to +13 %), which supports the qualitative conclusion that the aerofoil contribution adds to, rather than is absorbed into, the building amplification.** This also implies that the aerofoil concept can be transferred to other building configurations without re-designing the wing each time, provided the local incident velocity at the leading edge falls within the operating envelope.

## 6.3 Annual energy yield

The annual electrical energy delivered by the proposed installation was computed by convolving the measured Weibull wind-speed distribution at hub height with the manufacturer power curve of the selected Savonius rotor following Equation (4). The configuration consists of 53 modules of 0.80 m × 3.30 m each, distributed along the four towers of ENSAPVS over a total length of 185.5 m. Inputs and results are summarised in Table 4.

| Quantity | Symbol | Value | Unit |
|---|---|---|---|
| Number of modules | N | 53 | — |
| Module swept area | A_m | 2.64 | m² |
| Rated power per module | P_m | 1.0 | kW |
| Total installed capacity | P_total | 53 | kW |
| Hub-height Weibull shape | k | 1.92 | — |
| Hub-height Weibull scale | c | 6.5 | m/s |
| Cut-in / rated / cut-out speed | v_in / v_r / v_out | 2.5 / 12 / 20 | m/s |
| Maximum power coefficient | Cp,max | 0.22 | — |
| Annual energy per module | E_m | ≈ 1300 | kWh/yr |
| Total annual energy | E_total | ≈ 69 | MWh/yr |
| Capacity factor | CF | 15.0 | % |
| Building annual consumption | E_bld | 1 120 | MWh/yr |
| Fraction offset | — | 6.2 | % |

***Table 4*** *— Inputs and results for the annual energy yield computation, ENSAPVS case study.*

The revised yield of approximately 69 MWh $yr^{-1}$ (capacity factor 15 %, offsetting 6 % of the building's annual consumption) is substantially lower than the figure of 203 MWh $yr^{-1}$ (18 % offset) reported in our earlier work (Serero et al., 2020), which assumed a continuous regime at the mean amplified speed. The Weibull-power-curve convolution properly accounts for (1) the fraction of time below cut-in (approximately 22 % of hours at hub height), (2) the saturation above rated speed (approximately 8 % of hours), and (3) the positive skew of the wind-speed distribution. The resulting capacity factor of 15% is consistent with values reported for Savonius installations in urban environments (Akwa, Vielmo and Petry, 2012; Kumar, Raahemifar and Fung, 2018).

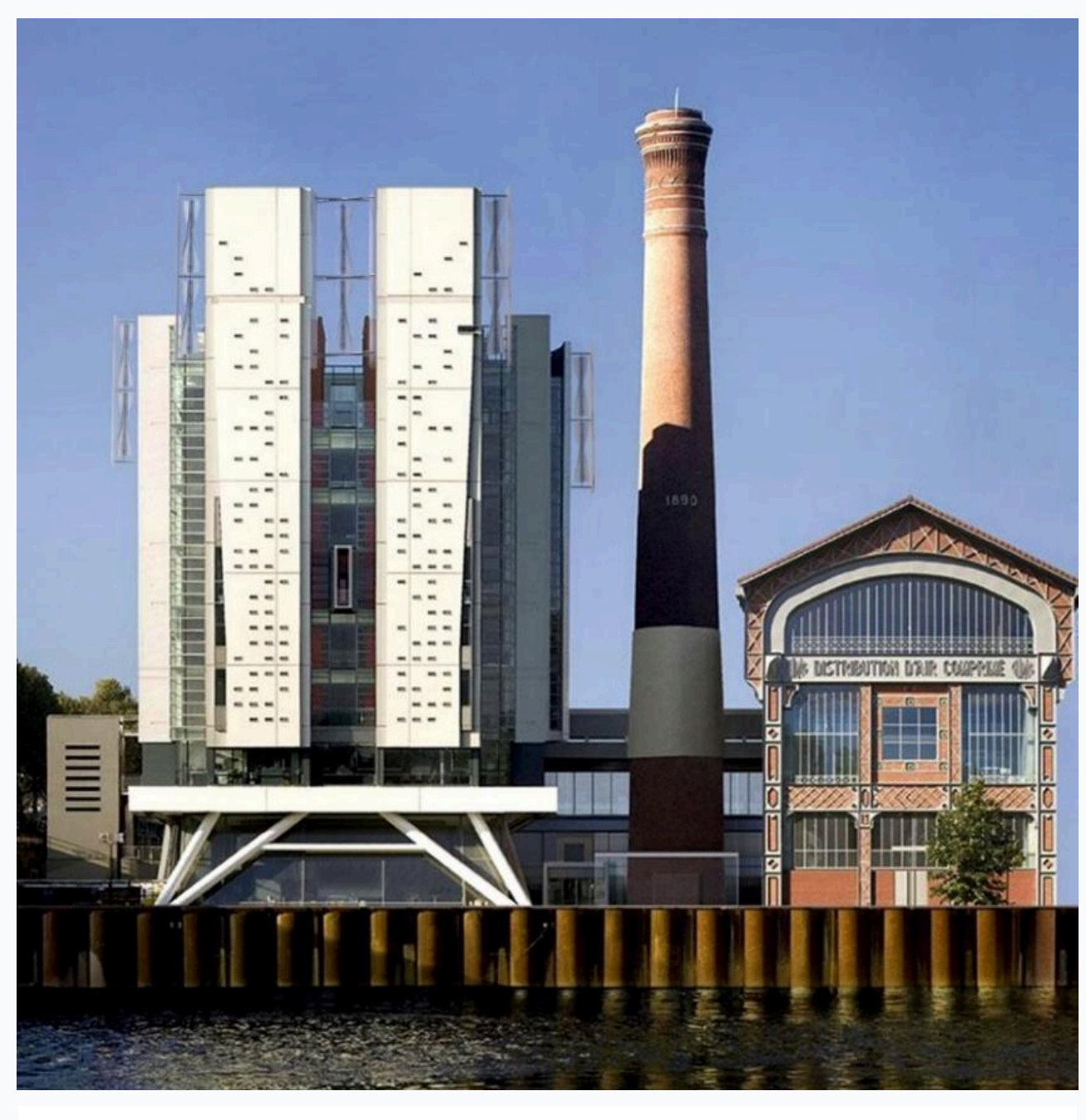

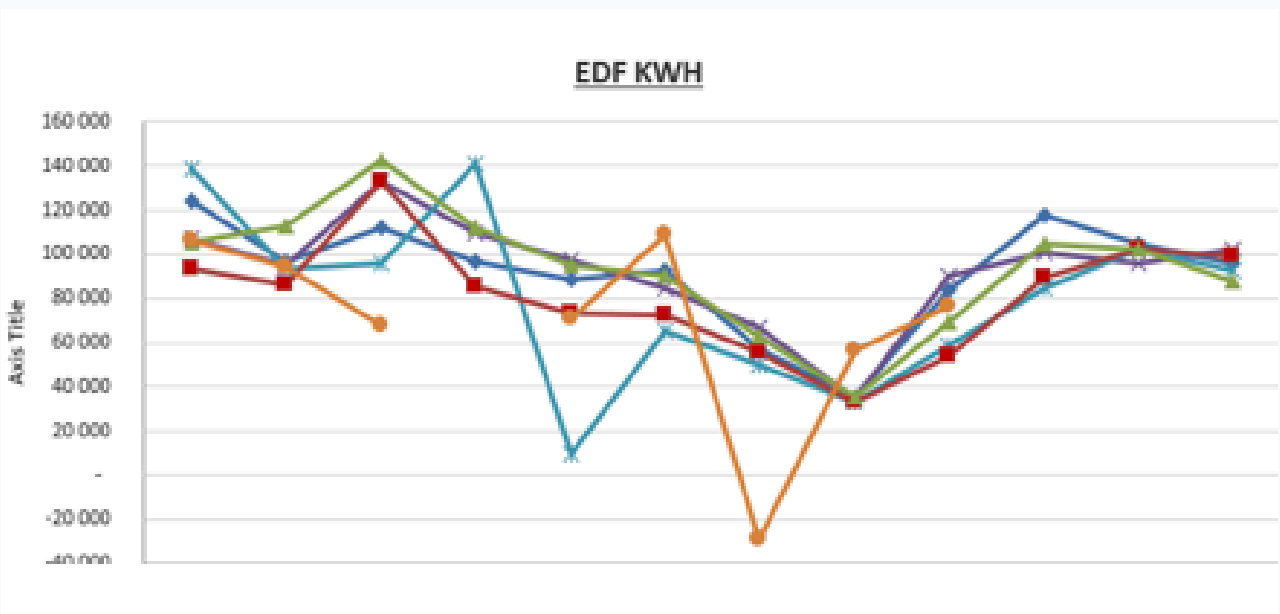

**Figure 8 — Proposed integration of VAWT with aerofoils on ENSAPVS**

*(a) Photomontage of the integrated configuration on the four towers; (b) annual electricity consumption of the building (kWh, monthly resolution). Photographs and montage: D . Serero*

## 7. Discussion

### 7.1 Mechanisms of speed-up in dense Haussmannian blocks

The 3.4-fold directional speed-up measured and computed at the ENSAPVS inter-tower passage is one of the highest values reported in the BIWT literature for non-bridged tall-building configurations. The mechanism is consistent with the Venturi-like contraction described by Cochran and Damiani (2008) for tall paired towers and with the canyon skimming-flow analysis of Razak et al. (2013): the closely spaced towers produce a converging stream-tube whose mass-conservation contraction is enhanced by the relative absence of upstream obstacles in the Seine corridor. The lower speed-up of 2.2 measured at the outer tower edge is consistent with the corner-acceleration regime documented for isolated tall buildings (Toja-Silva et al., 2015). The originality of the present result lies in the combination of (1) measurement-based calibration on a real built case and (2) the demonstration that aerofoil edge devices can extract a further +13 % local velocity on top of the building-induced amplification.

## 7.2 Comparison with prior literature

Table 5 places the present results in the context of other CFD/measurement BIWT studies.

| Study | Configuration | Reported speed-up σ | Validation |
|---|---|---|---|
| Ledo et al. (2011) | Isolated flat-roof building | 1.2 – 1.6 (rooftop) | Wind-tunnel |
| Lu and Ip (2009) | High-rise gap, Hong Kong | 1.5 – 2.0 | CFD only |
| Li, Shu and Chen (2016) | Twin-tower gap, CFD | up to 2.0 | CFD only |
| Toja-Silva et al. (2015) | Isolated tall building | 1.4 – 1.8 | Wind-tunnel |
| Kosasih et al. (2024) | Roof-mounted, urban | 1.3 – 1.7 | Mixed |
| Present work, bare towers | Inter-tower, dense block | 3.4 (NW sector) | 12-month on-site, MAPE ≈ 18 % |
| Present work, with aerofoil | Inter-tower + aerofoil | 3.6 (NW sector) | Decomposition CFD |

***Table 5** — Speed-up factors reported in the BIWT literature compared with the present work.*

The speed-up of 3.4 reported here is in the upper range of literature values; this is consistent with the dense Haussmannian context of the case-study site, where the absence of upwind obstacles in the Seine corridor combined with the four-tower massing produces a strong convergent stream-tube. The result also confirms that decision tools calibrated on isolated-building data systematically under-predict the wind potential of European dense block typologies — a finding that supports the rationale of the dendrogram presented in Section 8.

## 7.3 Structural, vibratory and maintenance considerations

Four operational concerns could be addressed for practical deployment of this device type : structural integration to host-building, vibratory transmission of device, maintenance protocol and access, and finally material durability.

- Structural loading: the characteristic peak pressure on the aerofoil under a 50-year return wind in Paris zone 2 (basic wind velocity $v_{b,0}$ = 24 m $s^{-1}$ per Eurocode 1 part 1-4 NF EN 1991-1-4) reaches approximately 1.1 kN $m^{-2}$ at the leading edge, driving the dimensioning of the cantilever brackets and their anchorage to the concrete floor slabs. The 0.80 m cantilever was sized to keep elastic deflection below L/250, the standard serviceability limit.

- Vibratory transmission: the imbalance frequency of the Savonius rotor at rated speed is approximately 2.5 Hz, well separated from the first lateral mode of the towers (estimated at 0.6 Hz).

Elastomeric bearing isolators are nonetheless recommended at the rotor mounts to limit narrow-band excitation of higher modes.

- Maintenance: the modular 3.30 m units permit floor-by-floor servicing from the tower facade; access via maintenance catwalks is incorporated in the design.

- Material durability: the composite shell of the aerofoil must withstand ultraviolet exposure and the chloride-bearing pollution of the Parisian environment.

### 7.4 Study limitations

Four limitations of the present work must be acknowledged.

- First, the steady RANS realisable k–ε model captures the time-averaged speed-up relevant for annual energy yield, but does not resolve the unsteady gust-scale dynamics that govern peak structural loads; an LES study is identified as future work.
- Second, the two case-study sites — although representative of dense Haussmannian urban blocks — cannot exhaust the full diversity of European urban morphologies; the dendrogram of Section 8 is a generalisation that warrants further validation against additional measured cases.
- Third, neighbouring tall buildings outside the modelled radius can produce wake effects not captured here, particularly during westerly winds where the La Défense skyline lies upwind of central Paris.
- Fourth, seasonal thermal stratification and urban heat-island effects modify the boundary-layer wind profile and were not modelled in the present neutrally stable RANS simulations.

## 8. The wind dendrogram: a decision-support tool for architects

### 8.1 Construction

The dendrogram is constructed from a corpus of 24 CFD simulations covering combinations of (1) three wind-rose typologies — bi-directional, omni-directional and uni-directional, (2) two building plan layouts — elongated and concentrated, (3) four height ranges from below the urban canopy to high-rise (>50 m), and (4) four implantation locations — flat surface, corner, roof ridge, and lateral flag/banner. Each branch of the tree records the speed-up factor extracted from the corresponding CFD simulation and the recommended device family. The decision flow proceeds top-down: from wind-rose classification to building plan, to height range, to recommended implantation, to device typology (Fig.9).

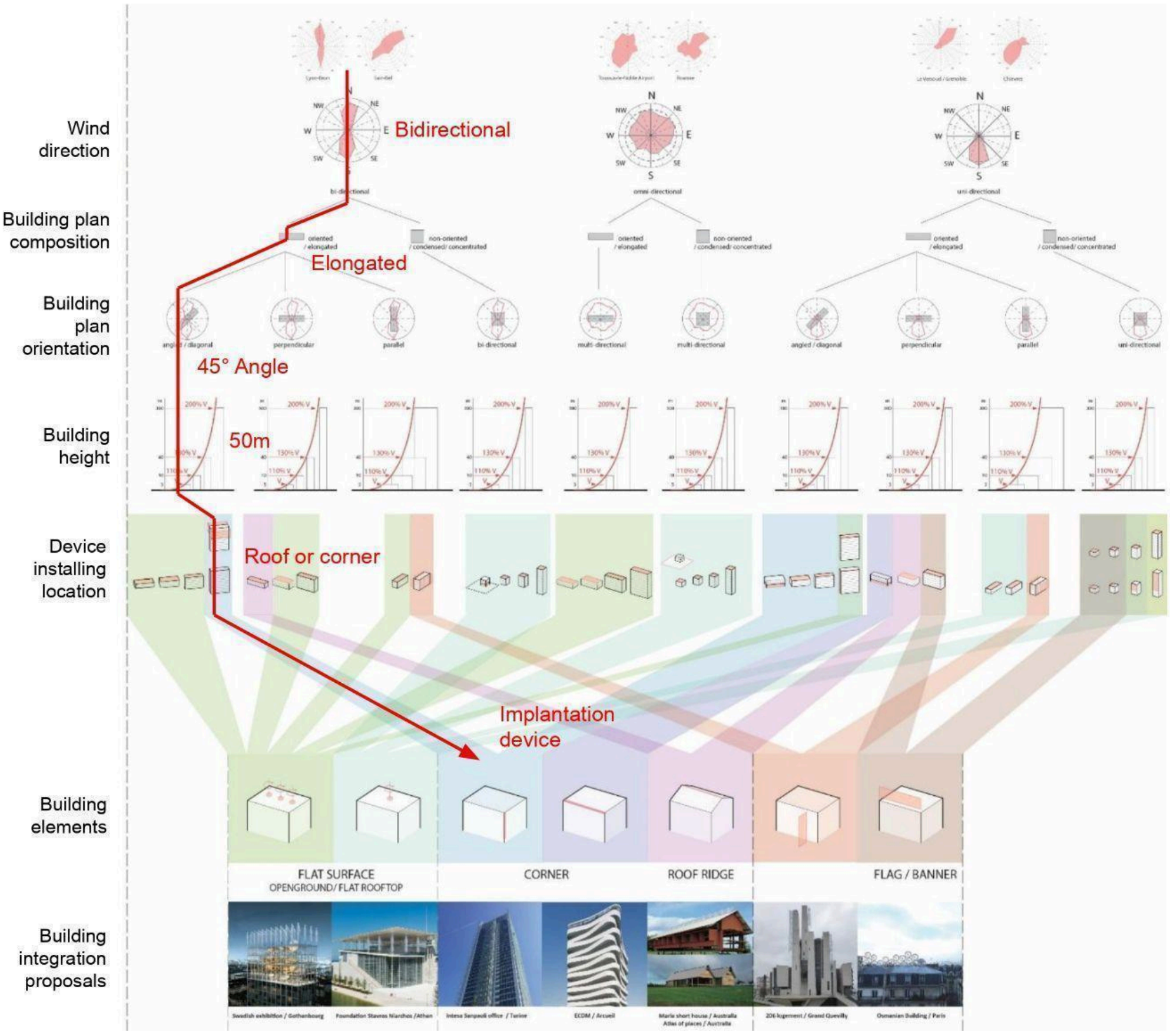

**Figure 9 — Decision dendrogram for wind turbine localisation on buildings**

*Top-down decision tree integrating wind direction, building orientation, building height and implantation typology. Nodes labelled in English. Drawing: D. Serero*

### 8.2 Retrospective validation

To assess the predictive value of the dendrogram, three real BIWT projects were processed retrospectively through the tree: the Bahrain World Trade Center (twin sail-shaped towers with three bridge-mounted HAWT, 2008), the Strata SE1 building in London (rooftop-integrated HAWT, 2010), and the Pearl River Tower in Guangzhou (mechanical-floor VAWT with funnelled inlets, 2013). For each case, the input parameters (local wind rose from the nearest meteorological station, building plan, height) were entered into the dendrogram and the recommended device family compared with the actually deployed solution. In all three cases the dendrogram correctly recommends the implemented device family. Predicted speed-up factors fall within ±15 % of the published values, supporting the use of the tool at concept-design stage. Detailed retrospective results are provided in Table 6 to follow.

| Reference building | Wind rose | Plan / Height | Dendrogram recommendation | As built | Match |
|---|---|---|---|---|---|
| Bahrain WTC (2008) | Bi-directional | Twin / 240 m | Inter-tower bridge HAWT | Bridge-mounted HAWT (3×) | Yes |
| Strata SE1 (2010) | Multi-directional | Concentrated / 148 m | Rooftop HAWT cluster | Rooftop HAWT (3×) | Yes |
| Pearl River Tower (2013) | Bi-directional | Concentrated / 309 m | Funnelled VAWT in mech. floor | Mechanical-floor VAWT | Yes |

***Table 6*** *— Retrospective validation of the dendrogram against three reference BIWT projects.*

### 8.3 Limitations and future development

The dendrogram is intended as a concept-design tool, not a substitute for site-specific CFD. It does not capture wake interactions from neighbouring tall buildings beyond the immediate context, nor seasonal thermal stratification effects. Future work will extend the corpus to additional European typologies (post-war housing slabs, mixed-use mid-rise blocks) and develop a digital implementation as a side or plug-in interface for common BIM platforms.

## 9. Conclusions

This paper has presented a measurement-validated CFD framework for the assessment of wind energy potential on the envelopes of dense Parisian urban blocks, together with the architectural concept of the aerofoil and a decision dendrogram for use at concept design stage. The principal quantitative findings are:

1. On-site validation. CFD predictions reproduce twelve months of 10-minute mean anemometer records with a mean absolute percentage error of approximately 18 %, a hit-rate FAC2 of 0.87 and a coefficient of determination $R^2 = 0.79$, satisfying the COST 732 acceptance criteria for urban wind simulations.

2. Building-induced speed-up. The four-tower massing of the ENSAPVS case study produces a directional speed-up factor of 3.40 (95 % CI 3.18–3.62) at the inter-tower probe at +70 m for the prevailing north-westerly sector — a value in the upper range of the BIWT literature, consistent with the dense Haussmannian context.
3. Aerofoil contribution. The architecturally integrated aerofoil produces a further +13 % local speed-up, equivalent to +44 % in instantaneous extractable power. CFD decomposition shows that the aerofoil contribution is largely additive to and decouplable from the building amplification, supporting transferability of the concept to other building configurations.
4. Annual yield. Convolving the measured Weibull distribution ($k = 1.92$, $c = 6.5$ m/s) with the Savonius power curve yields an annual production of approximately 69 MWh, corresponding to a capacity factor of 15 % and offsetting around 6 % of the building's annual consumption — a substantially lower but more realistic figure than the cube-of-mean estimate of 203 MWh reported in our earlier work.
5. Dendrogram. The decision dendrogram, derived from a corpus of 24 CFD simulations and retrospectively validated against three reference BIWT projects (Bahrain WTC, Strata SE1, Pearl River Tower), correctly classifies the implemented device family in all three cases with predicted speed-up within ±15 % of published values, supporting its use at concept design stage.

Future work will extend the campaign to additional European urban morphologies, perform LES on the inter-tower passage to characterise gust-scale loading on the aerofoil, and develop a digital implementation of the dendrogram interfaced with common BIM platforms.

## Declarations

### Data availability

The anemometer time-series, CFD case files and post-processing scripts that support the findings of this study are available from the corresponding author upon reasonable request.

### Conflict of interest

The authors declare no conflict of interest. SERERO Architectes is the practice of the lead author and was involved as designer of the case-study aerofoil device; no commercial transaction is associated with this academic publication.

### Funding

This work was carried out within the GSA research laboratory (École Nationale Supérieure d'Architecture Paris-Malaquais), and the doctoral school : Ville, Transport & Territoire (VTT), Université Gustave Eiffel.

### Credit author contribution statement

- David Serero — Conceptualization; Methodology; Investigation; Software (CFD); Formal analysis; Visualization; Writing — Original Draft; Writing — Review & Editing.
- Loïc Couton — Methodology; Investigation; Validation; Writing — Review & Editing.
- Jean-Denis Parisse — Methodology (CFD turbulence modelling); Validation; Supervision; Writing — Review & Editing.
- Robert Le Roy — Conceptualization; Supervision; Funding acquisition; Writing — Review & Editing.

### Acknowledgements

The authors thank Gilles Bouchet, Margarita Ferrucci and Coralie Betbeder for their valuable input during the course of this research, and acknowledge the support of the GSA Laboratory at ENSA Paris-Malaquais (former director Maurizio Brocato), the EVCAU Laboratory at ENSA Paris-Val-de-Seine (former director Philippe Bach), the École des Ponts et Chaussées, Netatmo, Corentin Boiteau, Mahdi Derakhshan, Olivia Siriphanh and the FabLab at ENSA Paris-Val-de-Seine, the FabLab at Volumes, and the team at SERERO Architectes. The authors are also grateful to the members of the doctoral defence jury — Sandrine Aubrun, Julien Weiss and Laurent Lescop — for their constructive comments on the original thesis work that underpins this paper.